\documentclass[a4paper,11pt]{article}
\usepackage{pos}

\usepackage{amsmath}
\usepackage{color}
\usepackage{cancel}
\usepackage{ulem}
\usepackage{bm}% bold math
\usepackage{braket}

\newcommand{\bec}[1]{\mbox{\boldmath $#1$}}
\newcommand{\oos}{\Omega\Omega}
\newcommand{\ooc}{\Omega_{ccc}\Omega_{ccc}}

\title{Finite volume analysis on systematics of the derivative expansion in HAL QCD method}
\author*[a,b]{Takumi Doi}
\author[c,a]{Yan Lyu}
\author[c,b]{Hui Tong}
\author[b]{Takuya Sugiura}
\author[d,a]{Sinya Aoki}
\author[b]{Tetsuo Hatsuda}
\author[c,e]{Jie Meng}
\author[a]{Takaya Miyamoto}

\affiliation[a]{Quantum Hadron Physics Laboratory, RIKEN Nishina Center,\\
  Wako 351-0198, Japan}

\affiliation[b]{Interdisciplinary Theoretical and Mathematical Sciences Program (iTHEMS), RIKEN,\\
  Wako 351-0198, Japan}

\affiliation[c]{State Key Laboratory of Nuclear Physics and Technology, School of Physics, Peking University,\\
  Beijing 100871, China}

\affiliation[d]{Center for Gravitational Physics, Yukawa Institute for Theoretical Physics, Kyoto University,\\
  Kyoto 606-8502, Japan}

\affiliation[e]{
  Yukawa Institute for Theoretical Physics, Kyoto University,\\
  Kyoto 606-8502, Japan}

\emailAdd{doi@ribf.riken.jp}
\emailAdd{helvetia@pku.edu.cn}
\emailAdd{tong16@pku.edu.cn}
\emailAdd{takuya.sugiura@riken.jp}
\emailAdd{saoki@yukawa.kyoto-u.ac.jp}
\emailAdd{thatsuda@riken.jp}
\emailAdd{mengj@pku.edu.cn}
\emailAdd{miyamoto@ribf.riken.jp}

\abstract{
  We study the convergence of the derivative expansion
  in HAL QCD method from the finite volume analysis.
  Employing the (2+1)-flavor lattice QCD data obtained at nearly physical light quark masses
  $(m_\pi, m_K) \simeq (146, 525)$ MeV and the physical charm quark mass,
  we study two representative systems, $\Omega\Omega$ and $\Omega_{ccc}\Omega_{ccc}$ in the $^1S_0$ channel,
  where both systems were found to have a shallow bound state in our previous studies.
  The HAL QCD potentials are determined at the leading-order in the derivative expansion,
  from which finite-volume eigenmodes are obtained.
  Utilizing the eigenmode projection,
  we find that the correlation functions are dominated by
  the ground state (first excited state) in the case of $\Omega\Omega$ ($\Omega_{ccc}\Omega_{ccc}$).
  In both $\Omega\Omega$ and $\Omega_{ccc}\Omega_{ccc}$,
  the spectra obtained from eigenmode-projected temporal correlators
  are found to be consistent with those from the HAL QCD potential for both
  the ground and first excited state.
  These results show that the derivative expansion is well converged in these systems,
  and also provide a first explicit evidence that
  the HAL QCD method enables us to reliably extract
  the binding energy of the ground state
  even from the correlator dominated by excited scattering states.
}

\FullConference{%
 The 38th International Symposium on Lattice Field Theory, LATTICE2021
  26th-30th July, 2021
  Zoom/Gather@Massachusetts Institute of Technology
}

\begin{document}
\begin{flushright}
  RIKEN-QHP-513,
  RIKEN-iTHEMS-Report-21,
  YITP-21-148
\end{flushright}
\maketitle

%%%%%%%%%%%%%%%%%%%%%%%%%%%%%%%%%%%%%%%%%%%%%%%%%%%%%%%%%%%%%%%%%%%%%%%%%%%%%%
\section{Introduction}
\label{sec:intro}

The determination of hadron interactions
is one of the most active areas in lattice QCD
these days.
Currently, there are two major theoretical methods,
L\"uscher's finite volume method~\cite{Luscher:1990ux}
and HAL QCD method~\cite{Ishii:2006ec, Ishii:2012ssm, Aoki:2020bew}.
The former first determines a finite volume spectrum
from a temporal correlation function,
and converts the spectrum to a scattering phase shift at that energy
through L\"uscher's formula.
The latter calculates
an energy-independent non-local potential
from a tempo-spatial correlation function,
and binding energies and phase shifts
are obtained by solving Schr\"{o}dinger-type equation
with the potential in the infinite volume.

While these two methods are equivalent theoretically,
each method has its own pros and cons in practical calculations.
In L\"uscher's method, it is essential to isolate each eigenstate
in the correlator, 
but it becomes difficult to suppress contaminations from nearby states
if the energy splittings between states are small.
In fact, we pointed out~\cite{Iritani:2018vfn} that naive plateau identification
for two-baron systems employed in the literature (so-called ``direct method'')
leads to unreliable results
due to the excited state contaminations.
Our finding is also being confirmed by recent
lattice QCD studies with L\"uscher's method~\cite{Francis:2018qch, Horz:2020zvv, Amarasinghe:2021lqa}.
On the other hand, HAL QCD method is free from such a problem
(as far as elastic states are concerned),
since one can extract the signal of energy-independent potential
even from excited states through its time-dependent formalism~\cite{Ishii:2012ssm}.
The method, however, introduces another type of systematic errors in practice,
because the non-locality of the potential is usually determined order by order
in the derivative expansion.
In Ref.~\cite{HALQCD:2018gyl}
we explicitly determined the potential up to next-to-next-to-leading order (N$^2$LO)
and found that the truncation error is well under control
even at the leading-order (LO)  at low energies.

Since the calculation of higher order terms in the derivative expansion
is usually expensive, we consider an alternative method~\cite{Iritani:2018vfn} in this report
to study the convergence of the derivative expansion.
The method essentially examines the consistency between L\"uscher's finite volume method
and the HAL QCD method, utilizing the finite volume eigenmodes obtained from the HAL QCD potential.
Considering that the origin of systematic errors of two methods are quite independent,
the observation of the consistency provides non-trivial confirmation that
the systematic errors are well under control.

In this report, we study two representative systems,
$\oos$ and $\ooc$ systems in the $^1S_0$ channel
near physical quark masses on a large volume ($La \simeq 8.1$ fm).
These systems have several good characteristics for our study:
(1) Heavy mass and large volume make
energy splittings between finite volume eigenstates small ($\sim$ a few MeV)
(2) Absence of valence ud-quarks makes statistical fluctuations small
(3) It was found that there is one shallow bound state for
each system~\cite{Gongyo:2017fjb, Lyu:2021qsh}.
In addition, it turns out that
the correlator of each system serves as two representative cases, i.e.,
the $\oos$ correlator is dominated by the ground state,
while
the $\ooc$ correlator is dominated by the first excited state.
In this report, we examine whether
binding energies and scattering phase shifts are reliably extracted
for both cases in the HAL QCD method.

%%%%%%%%%%%%%%%%%%%%%%%%%%%%%%%%%%%%%%%%%%%%%%%%%%%%%%%%%%%%%%%%%%%%%%%%%%%%%%
\section{HAL QCD method and finite volume analysis}
\label{sec:method}

The key quantity in the HAL QCD method
is the equal-time Nambu-Bethe-Salpeter (NBS) wave function.
In the case of a $BB$ system with $B=\Omega$ or $\Omega_{ccc}$ in this study,
it is defined by
%
%
%\begin{eqnarray}
$
\phi_W(\bec{r}) \equiv 
1/Z_{B} \cdot
\langle 0 | \hat{B}(\bec{r},0) \hat{B}(\bec{0},0) | BB, W \rangle,
$
%\end{eqnarray}
%
where 
$\hat{B}$ is an operator for a $B$-baryon
with its wave-function renormalization constant $Z_B$ 
and
$|BB, W \rangle$ denotes the $BB$ eigenstate
at the total energy of $W = 2\sqrt{k^2+m_B^2}$,
and we consider the elastic region, $W < W_{\rm th}$.
%
%
%
%The most important property of the NBS wave function is that
%the information of the phase shift $\delta_l(k)$ ($l$: the orbital angular momentum)
%is encoded in the asymptotic behavior
%at $r \equiv |\bec{r}| \rightarrow \infty$
%as
%$
%\phi_W (\bec{r}) \propto 
%\sin(kr-l\pi/2 + \delta_l(k)) / (kr).
%$
%
%
%Exploiting this feature,
Since the information of the phase shift is encoded in the asymptotic behavior
($r \equiv |\bec{r}| \rightarrow \infty$)
of the NBS wave function,
one can define energy-independent non-local potential, $U(\bec{r},\bec{r}')$,
%which is faithful to the phase shifts
through the Schr\"odinger equation,
%
%\begin{eqnarray}
%
$
(E_W - H_0) \phi_W(\bec{r})
= 
\int d\bec{r'} U(\bec{r},\bec{r'}) \phi_W(\bec{r'}) ,
$
%\label{eq:Sch_2N:tindep}
%\end{eqnarray}
%
where 
$H_0 = -\nabla^2/(2\mu)$ and
$E_W = k^2/(2\mu)$ with the reduced mass $\mu = m_B/2$~\cite{Ishii:2006ec, Ishii:2012ssm, Aoki:2020bew}.

Generally speaking, the NBS wave function at each eigenenergy 
can be extracted from the four-point correlator.
Such a procedure, however, is exponentially difficult in practice,
if one relies on the ground state (or eigenstate) saturation
utilizing the temporal behavior of the correlator.
The time-dependent HAL QCD method~\cite{Ishii:2012ssm} overcomes this problem
by exploiting the fact that the same potential governs all elastic states.
Namely,
we define the normalized four-point function %(the $R$ correlator)
as
$
R(\bm{r},t) \equiv \sum_{\bm x} \langle 0|\hat{B}(\bm{r} + \bm{x},t)\hat{B}(\bm{x},t)
\overline{\mathcal{J}}(0)|0\rangle/e^{-2m_Bt}
$
and
the potential can be determined by the following master formula:
\begin{equation}
  \left(\frac{1}{4m_B}\frac{\partial^2}{\partial t^2}-\frac{\partial}{\partial t}-H_0\right) R(\bm{r},t) = \int d\bm{r}'U(\bm{r},\bm{r}')R(\bm{r}',t) .
\end{equation}
The systematic error in this equation is
the contaminations from the inelastic states,
which can be suppressed by taking moderately large Euclidean time, $t \gg (W_{\rm th} - W)^{-1}$.

In practical calculation, non-locality of the potential is handled by
the derivative expansion at low energies,
%$U(\bm{r},\bm{r}')=V(r)\delta(\bm{r}-\bm{r}')+ \sum\limits_{n=1}V_{2n}(\bm{r})\nabla^{2n}\delta(\bm{r}-\bm{r}')$.
$U(\bm{r},\bm{r}')= \sum_{n} V_{n}(\bm{r})\nabla^{n}\delta(\bm{r}-\bm{r}')$.
For instance, the central potential $V(r)$ at the LO is given as
\begin{equation}
  V(r)=R^{-1}(\bm{r},t)\left(\frac{1}{4m_B}\frac{\partial^2}{\partial t^2}-\frac{\partial}{\partial t}-H_0\right) R(\bm{r},t).
\end{equation}
This procedure, however, introduces new systematic errors
associated with the truncation in the derivative expansion.
In order to quantify such systematic uncertainties,
it is most desirable to calculate the higher order terms explicitly and
examine the convergence.
Such a study was performed up to N$^2$LO for the $\Xi\Xi$ system 
in the $^1S_0$ channel at $m_\pi = 0.51$ GeV~\cite{HALQCD:2018gyl},
and it was found that the truncation error is well under control
even at the LO  at low energies.
However, the explicit computations of higher order terms require large resources,
which leads us to study an alternative method in this report.

We here perform the finite volume analysis~\cite{Iritani:2018vfn}
as a new measure which can quantify the systematic uncertainties
of the derivative expansion without requiring any additional lattice QCD simulation.
In this analysis, we first consider the following 
Hamiltonian $H$ in a finite box,
and calculate its eigenenergies and eigenfunctions (eigenmodes),
\begin{equation}\label{Eq_H}
H=H_0 + V(r), \quad H\psi_n=\epsilon_n\psi_n,
\end{equation}
where $V(r)$ is the HAL QCD potential obtained at the LO in the derivative expansion,
and $\epsilon_n$ is related to the relativistic energy, 
$W_n=2\sqrt{\epsilon_n m_B+m_B^2}$.
Obtained eigenfunctions are useful to isolate the contribution from each eigenmode
in the correlator.
More specifically, we utilize the eigenfunctions to construct
a two-baryon sink operator optimized for each eigenstate on a finite volume,
\begin{equation}
  \sum_{\bm r}\psi_n^\dagger(\bm r)\left[\sum_{\bm x}\hat{B}(\bm r + \bm x, t)\hat{B}(\bm x, t)\right].
  \label{eq:opt-op}  
\end{equation}
A temporal correlator with such an optimized two-baryon sink operator can be obtained as follows,
\begin{equation}
  R_n(t) \equiv \sum_{\bm r}\psi^\dagger_n(\bm r)R(\bm r, t),
  \label{eq:opt-corr}
\end{equation}
where $\psi_n$ serves as a projection operator to the designated eigenstate.

One can extract the finite volume energy from this temporal correlator,
which value has one-to-one correspondence to the phase shift via L\"uscher's formula.
Note that, while the information of the HAL QCD potential is implicitly used to construct the
optimized operator, the theoretical formulation is solely based on L\"uscher's finite volume method.
The result of the finite volume spectrum can be compared
to that obtained directly from the HAL QCD potential (i.e., eigenenergies of $H$ on a finite volume).
If we observe consistency, it is non-trivial confirmation that
the systematic errors in the HAL QCD potential are well under control.

%%%%%%%%%%%%%%%%%%%%%%%%%%%%%%%%%%%%%%%%%%%%%%%%%%%%%%%%%%%%%%%%%%%%%%%%%%%%%%
\section{Lattice QCD setup}
\label{sec:setup}

Numerical data used in this study are obtained from the ($2+1$)-flavor gauge configurations
with Iwasaki gauge action at $\beta=1.82$ and nonperturbatively
$O(a)$-improved Wilson quark action with stout smearing
at nearly physical quark masses~\cite{Ishikawa:2015rho}.
The relativistic heavy quark (RHQ) action is used for the charm quark to remove cutoff errors
associated with the charm quark mass up to next-to-leading order,
with RHQ parameters determined in Ref.~\cite{Namekawa:2017nfw}.
The lattice cutoff is $a^{-1}\simeq2.333$~GeV ($a\simeq0.0846$~fm)
and the lattice volume is $(La)^4 = (96a)^4 \simeq (8.1\ {\rm fm})^4$.
The hadron masses most relevant to this study are 
$(m_\pi, m_K, m_{\Omega}, m_{\Omega_{ccc}}) \simeq (146, 525, 1712, 4796) $~MeV.
The NBS correlation functions
for $\oos$ and $\ooc$ systems %in the $^1S_0$ channel
are calculated
by the unified contraction algorithm~\cite{Doi:2012xd}
with the wall-type source operator combined with the Coulomb gauge fixing.
In order to reduce statistical fluctuations, forward and backward propagations are averaged,
the hypercubic symmetry on the lattice (4 rotations) are utilized,
and multiple measurements are performed by shifting the source position along the temporal direction.
The total measurements for $\oos$ ($\ooc$) amounts to 307200 (896).
For more details, see Refs.~\cite{Gongyo:2017fjb, Lyu:2021qsh}.

%%%%%%%%%%%%%%%%%%%%%%%%%%%%%%%%%%%%%%%%%%%%%%%%%%%%%%%%%%%%%%%%%%%%%%%%%%%%%%
\section{Results}
\label{sec:results}

\begin{figure}[t]
  \begin{minipage}{0.32\textwidth}
      \centering
      \includegraphics[width=0.95\textwidth]{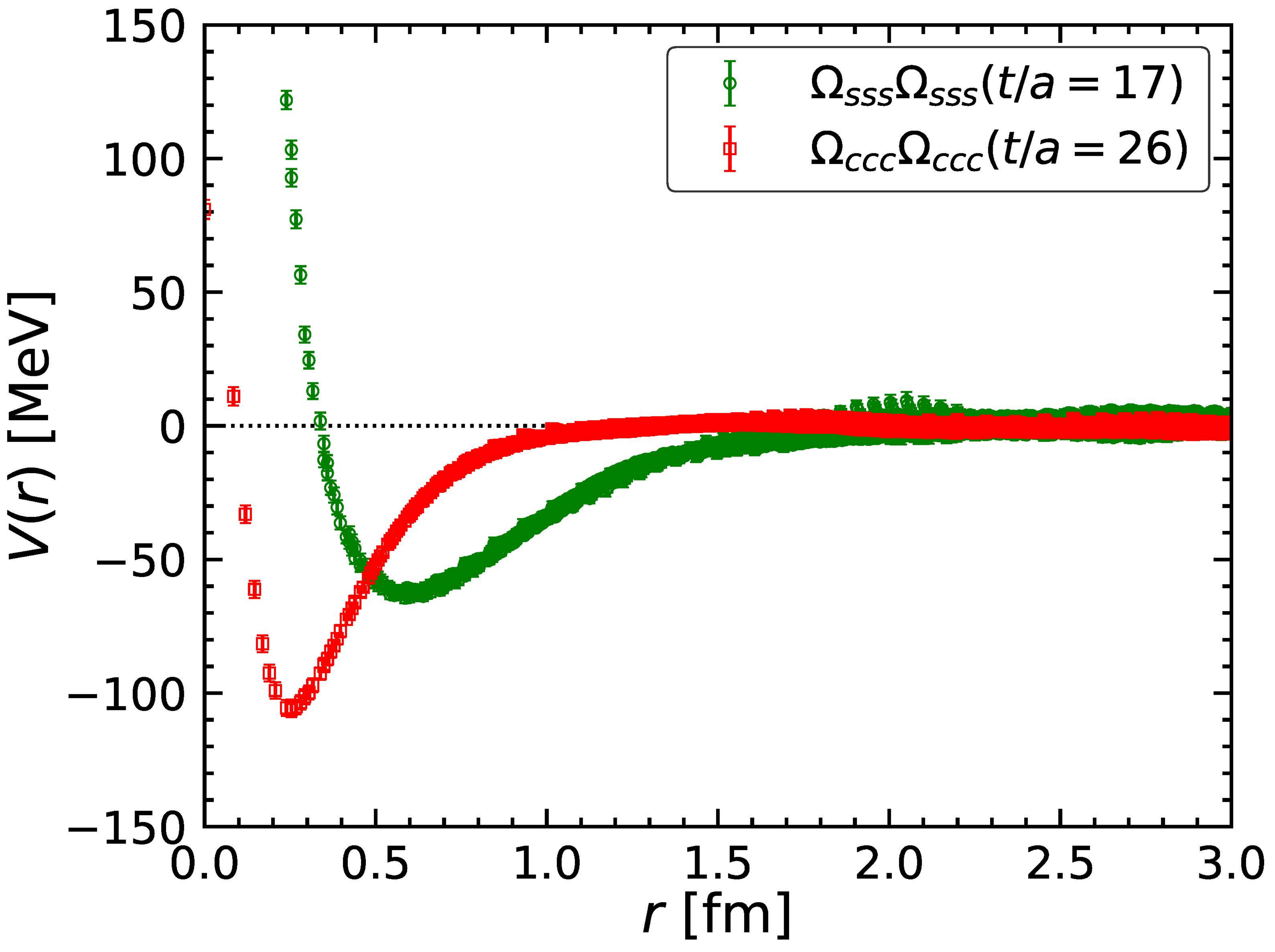}
      \caption{
        The LO HAL QCD potentials $V(r)$
        for $\oos(=\Omega_{sss}\Omega_{sss})$ (green circles) and $\ooc$ (red squares)
        in the ${^1S_0}$ channel.
      }
      \label{fig:pot}
  \end{minipage}
  \hfill
  \begin{minipage}{0.64\textwidth}
    \centering
    \includegraphics[width=0.48\textwidth]{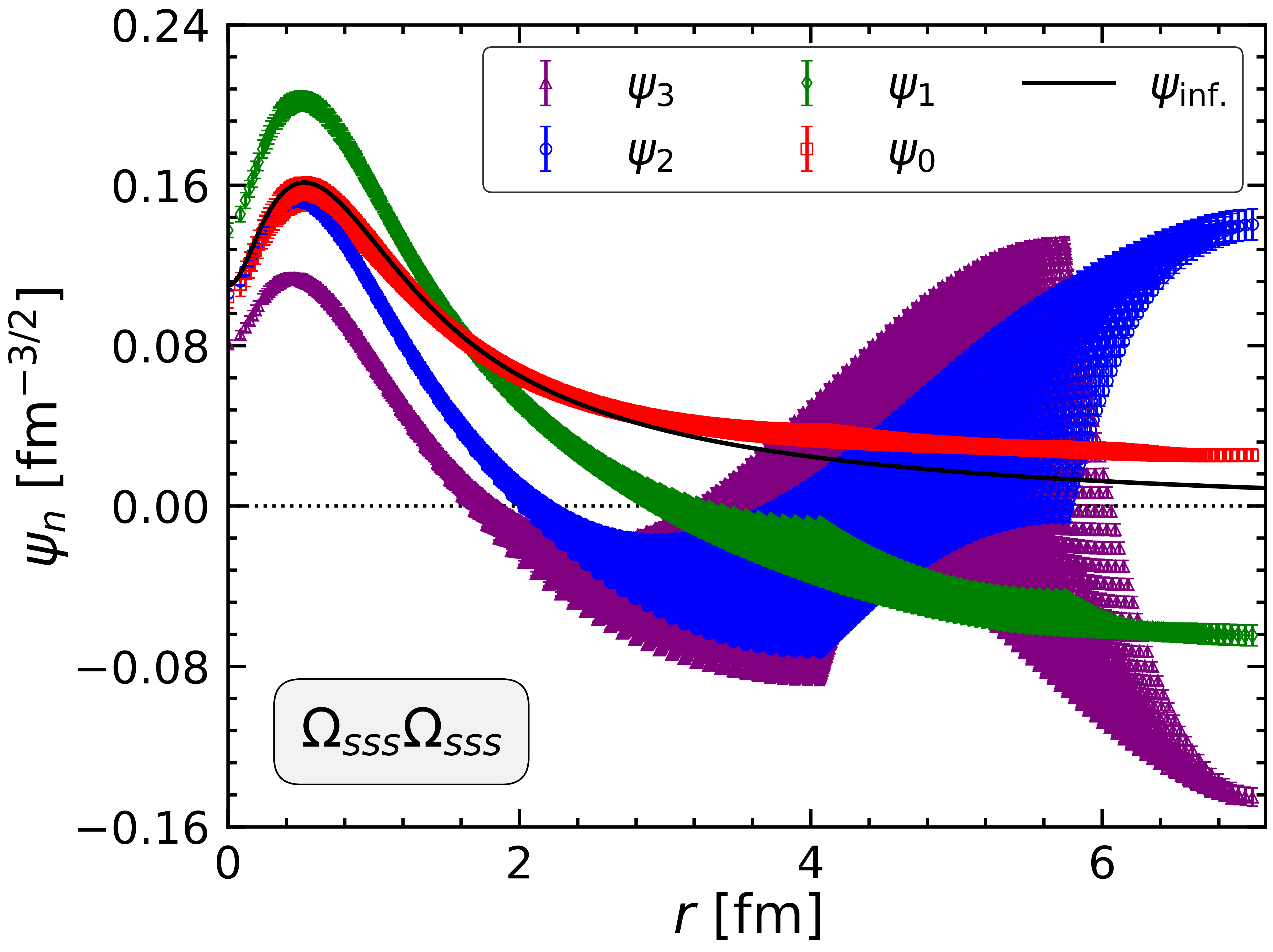}
    \includegraphics[width=0.48\textwidth]{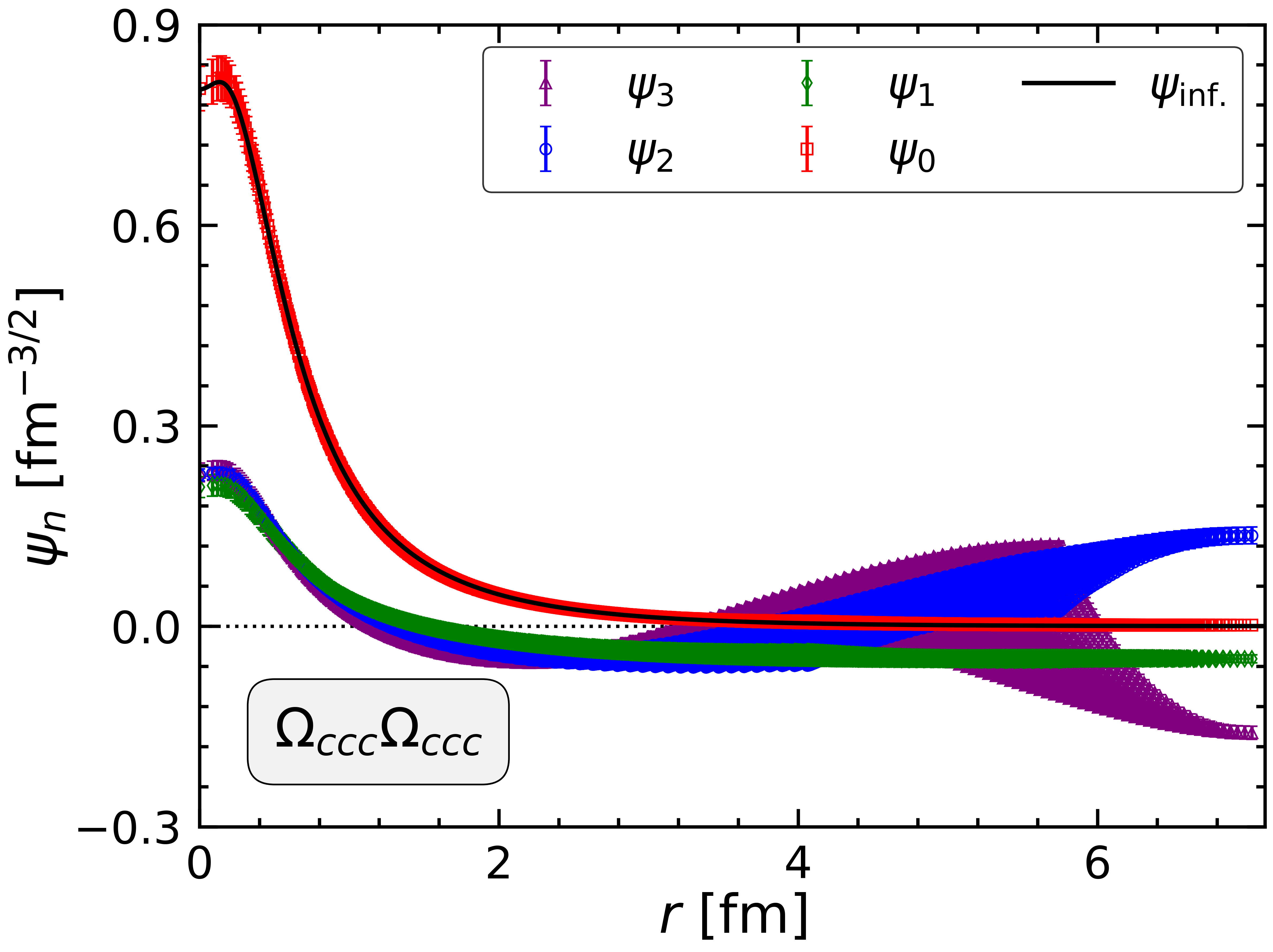}
    \caption{The lowest four eigenfunctions (colored points) in the $A_1$ representation of the HAL QCD Hamiltonian
      on a finite volume
      for $\oos(=\Omega_{sss}\Omega_{sss})$ (left) and $\ooc$ (right).
%      The red square, green diamond, blue circle, and purple triangle correspond to
%      the eigenfunctions of $n=0, 1, 2, 3$-th states, respectively.
      The black solid lines denote the bound state wavefunctions in the infinite volume.}
    \label{fig:wf}
  \end{minipage}
\end{figure}

In Fig.~\ref{fig:pot},
we show the LO HAL QCD potentials $V(r)$ in the ${^1S_0}$ channel for
$\oos$ at $t/a=17$
and
$\ooc$ at $t/a=26$,
where
$t/a$ are chosen
so that contaminations from inelastic excited states are suppressed
in the single-baryon correlators.
By solving the Schr\"odinger equation in the infinite volume,
we find that each system forms a loosely bound state
with the binding energy $B\simeq1.6$ MeV, root-mean-square distance $\sqrt{\langle r^2\rangle}\simeq3.4$~fm
for $\oos$~\cite{Gongyo:2017fjb}, and $B\simeq5.7$~MeV, $\sqrt{\langle r^2\rangle}\simeq1.1$~fm
for $\ooc$~\cite{Lyu:2021qsh}.
(If we consider the effect of the Coulomb repulsion,
both dibaryons are located in the unitary regime~\cite{Lyu:2021qsh},
but we study only QCD in this report.)

Using the obtained potentials,
we consider the HAL QCD Hamiltonian $H = H_0 + V$
on a finite volume %with periodic boundary condition.
and calculate eigenmodes in the $A_1$ representation. % of the cubic group $SO(3,\mathbb{Z})$.
The lowest four eigenfunctions  $\psi_n$ with $n=0, 1, 2$ and $3$ are shown in Fig.~\ref{fig:wf},
which are normalized as $\sum_{\bm r}|\psi_n(\bm r)|^2=1$ and $\psi_n(\bm 0)>0$.
Shown together are the bound state wavefunctions $\psi_\mathrm{inf.}$,
which are calculated in the infinite volume.

\begin{figure}[t]
  \begin{minipage}{0.48\textwidth}
    \centering
    \vspace*{-18mm}
    \includegraphics[width=0.95\textwidth]{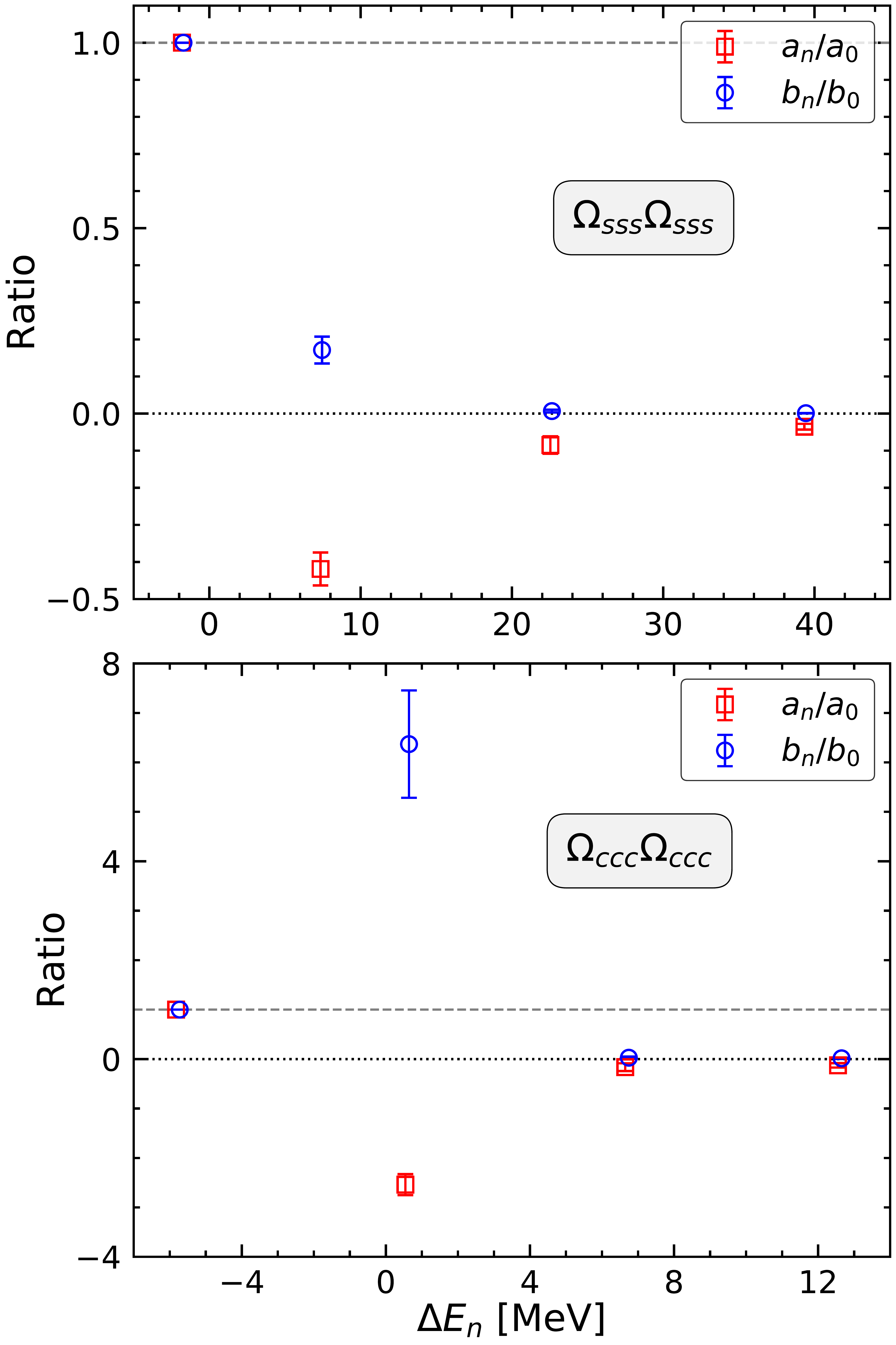}
    \caption{
      The ratio $a_n/a_0$ (red squares) and $b_n/b_0$ (blue circles)
      for lowest four states ($n=0,1,2,3$)
      as a function of $\Delta E_n$ for $\oos(=\Omega_{sss}\Omega_{sss})$ (upper) and $\ooc$ (lower).
    }
    \label{fig:ratio}
  \end{minipage}
  \hfill
  \begin{minipage}{0.48\textwidth}
    \centering
    \includegraphics[width=0.95\textwidth]{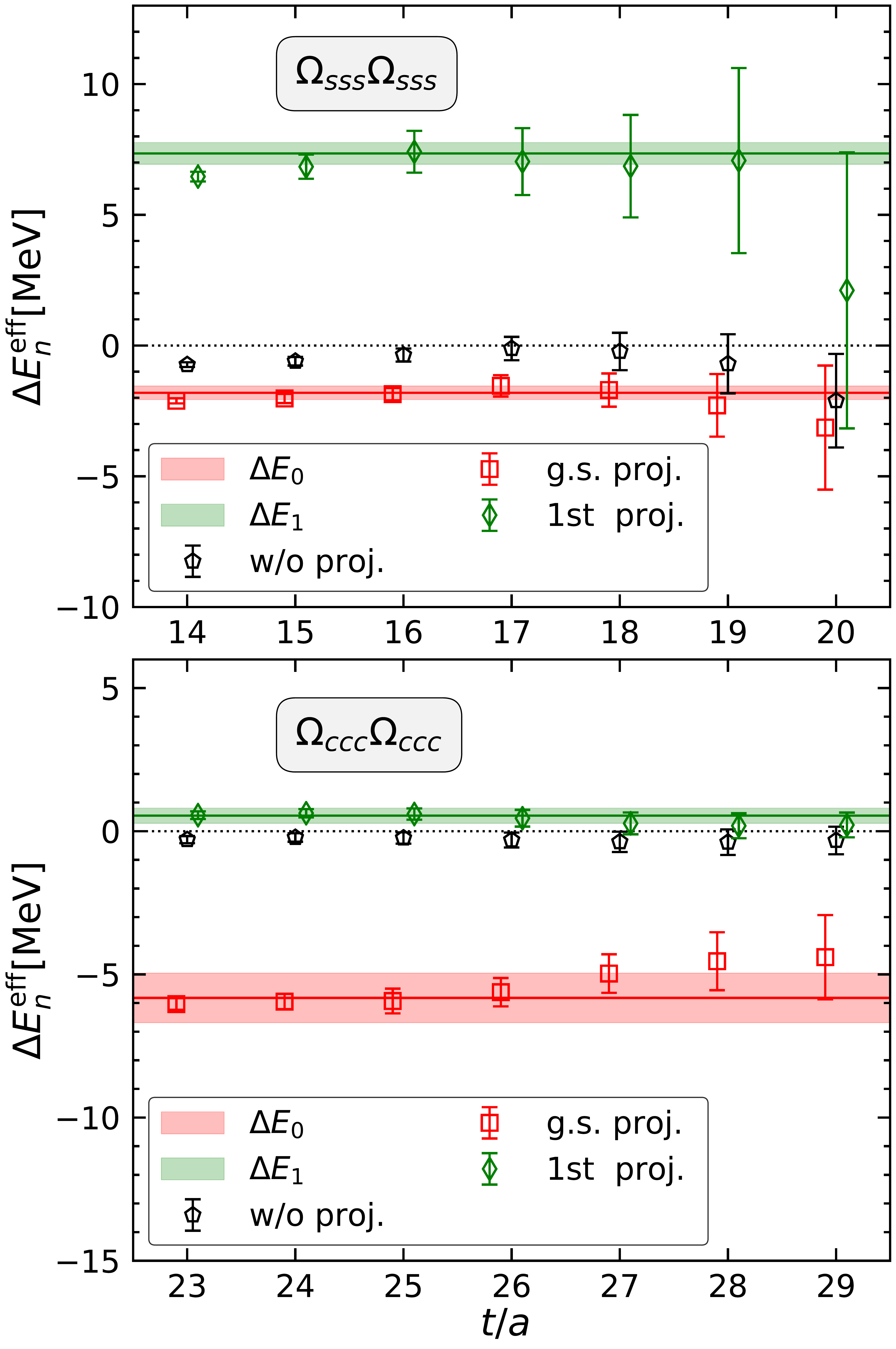}
    \caption{
      The effective energies $\Delta E^\mathrm{eff}_n(t)$
      from the projected temporal correlators $R_n(t)$
      for the ground state ($n=0$, red squares) and
      the first excited state ($n=1$, green diamonds)
      for $\oos$ (upper) and $\ooc$ (lower).
      The red (green) bands show $\Delta E_0$ ($\Delta E_1$) obtained from the HAL QCD Hamiltonian $H$.
      The black pentagons represent the effective energies from the temporal correlators without projection.
    }
    \label{fig:meff}
  \end{minipage}
\end{figure}

As described in Sec.~\ref{sec:method},
each eigenfunction can be used as the projection operator
for the corresponding eigenstate.
This enables us to decompose the correlators, $R(\bm r,t)$ and $R(t)$, as
\begin{equation}
 \begin{split}
    &R(\bm r,t)=\sum_n a_n\psi_n(\bm r)e^{-(\Delta E_n)t},\\
    &R(t)\equiv\sum_{\bm r}R(\bm r,t)=\sum_n b_ne^{-(\Delta E_n)t},
 \end{split}
\end{equation} 
with $\Delta E_n \equiv W_n - 2m_B$.
Here, $a_n$ and $b_n$ represent the magnitude of contribution of
the $n$-th state to the correlator $R(\bm r,t)$ and $R(t)$, respectively,
and they can be determined by
$a_n = \sum_{\bm{r}} \psi_n^\dag(\bm{r}) R(\bm r,t) e^{(\Delta E_n)t}$,
$b_n = a_n \sum_{\bm{r}} \psi_n(\bm{r})$
with $\Delta E_n$ calculated from the eigenenergies of $H$.
%in a finite volume.
%
%
In Fig.~\ref{fig:ratio},
we show $a_n/a_0$ and $b_n/b_0$ with $n=0, 1, 2$ and 3 for both of
$\oos$ and $\ooc$.
We find that two systems correspond to two different representative cases:
the correlator of $\oos$ is dominated by the ground state,
while that of $\ooc$ is dominated by the first excited state.
One can intuitively understand the origin of this observation.
In the case of $\oos$, the ground state wavefunction is long-ranged in $r$,
and thus the lattice source operator constructed by the wall source
is expected to couple the ground state strongly.
On the other hand, the ground state wavefunction of $\ooc$
is much more short-ranged, and the coupling between the lattice source operator
and the ground state is expected to be suppressed.

In order to examine the systematic uncertainties of the derivative expansion,
we construct the optimized two-baryon sink operator utilizing the eigenfunctions
(Eq.~(\ref{eq:opt-op})), and calculate the eigenmode-projected temporal correlator,
$R_n(t)$ (Eq.~(\ref{eq:opt-corr})).
In Fig.~\ref{fig:meff}, we show the effective energies $\Delta E^\mathrm{eff}_n(t)$
obtained from $R_n(t)$
with $n=0$ (the ground state) and $n=1$ (the first excited state) for both
$\oos$ and $\ooc$.
Shown together are the corresponding $\Delta E_n$
calculated directly from the HAL QCD Hamiltonian $H$.

In both cases of $\oos$ and $\ooc$,
we find that the effective energies $\Delta E^\mathrm{eff}_n(t)$
are stable in terms of $t$,
which values are consistent with $\Delta E_n$ for both of the ground state ($n=0$)
and the first excited state ($n=1$).
This establishes the consistency
between L\"uscher's finite volume method and
the HAL QCD method,
indicating that systematic errors associated with the truncation
of the derivative expansion are well under control.
In fact, if such artifacts were to be large, eigenfunctions $\psi_n(r)$
would be so different from the $n$-th eigenstate of the system that
the effective energies from $R_n(t)$ would  be distorted and do not agree with those from $H$.
Our observation also confirms that the HAL QCD method can make reliable predictions
regardless whether a correlator is dominated by the
ground state or (first) excited state.

This is in sharp contrast to the naive extraction of finite volume energies from
the temporal correlators.
To demonstrate this point, we show  in Fig.~\ref{fig:meff}
the effective energies from the temporal correlators without any sophisticated projection,
i.e., $R(t)=\sum_{\bm r}R(\bm r, t)$.
Such effective energies have been customary employed in the
``direct method'' to calculate the ground state spectrum for two-baryon systems.
We find that effective energies show stable plateau-like structures,
which values, however, significantly deviate from the correct values of the ground states.
In particular, the effective energy of $\ooc$ is very close to
the value of the first excited state,
reflecting the fact that the $\ooc$ correlator is dominated by the first excited state.
The calculations based on such pseudo-plateaux, of course, lead to unrealiable predictions.

%%%%%%%%%%%%%%%%%%%%%%%%%%%%%%%%%%%%%%%%%%%%%%%%%%%%%%%%%%%%%%%%%%%%%%%%%%%%%%
\section{Conclusions}
\label{sec:conclusion}

We studied the convergence of the derivative expansion
in the HAL QCD potential.
As good representative systems,
we considered $\oos$ and $\ooc$ in the $^1S_0$ channel,
each of which has a shallow bound state.
The lattice calculations were performed
in (2+1)-flavor QCD with nearly physical light quark masses
$(m_\pi, m_K) \simeq (146, 525)$ MeV and the physical charm quark mass.

The HAL QCD potentials were determined
at the leading-order in the derivative expansion.
The corresponding finite-volume eigenmodes were obtained,
from which the eigenmode projection on the correlator was performed.
Our finite volume analysis showed 
that the correlation function is dominated by
the ground state (first excited state) in the case of $\oos$ ($\ooc$).
For both $\oos$ and $\ooc$ systems,
the spectra obtained from eigenmode-projected temporal correlators
were found to be consistent with those from the HAL QCD potential.
This serves as a non-trivial consistency check between
L\"uscher's finite volume method
and
the HAL QCD method,
and confirms  that the derivative expansion in the HAL QCD method is well converged in these systems.
In addition, we conclude
that 
the HAL QCD method can make reliable predictions
regardless whether a correlator is dominated by the
ground state or (first) excited state.
It is particularly striking that
the HAL QCD potential obtained from
the correlator dominated by the first excited state
can determine the binding energy of the ground state reliably.

%%%%%%%%%%%%%%%%%%%%%%%%%%%%%%%%%%%%%%%%%%%%%%%%%%%%
%\vspace*{-1mm}
\section*{Acknowledgments}
%\vspace*{-1mm}

We thank the members of HAL QCD Collaboration, the members of PACS Collaboration,
Yusuke Namekawa, Tatsumi Aoyama, Ken-Ichi Ishikawa, Haozhao Liang, Shuangquan Zhang and Pengwei Zhao
for their supports and stimulating discussions.
We thank ILDG/JLDG~\cite{Amagasa:2015zwb} and the authors of
domain-decomposed solver~\cite{dd-solver},
cuLGT code~\cite{Schrock:2012fj}
and Bridge++ code~\cite{bridge}.
The lattice QCD calculations have been performed on
K computer at RIKEN,
HOKUSAI supercomputers at RIKEN
and
HA-PACS at University of Tsukuba.
This work was partially supported by HPCI System Research Project
(hp120281, hp130023, hp140209, hp150223, hp150262, hp160211, hp170230, hp170170, hp180117, hp190103,
hp200130, hp210165),
the National Key R$\&$D Program of China (2017YFE0116700, 2018YFA0404400),
the National Natural Science Foundation of China (11935003, 11975031, 11875075, 12070131001),
JSPS Grant (JP18H05236, JP16H03978, JP19K03879, JP18H05407),
MOST-RIKEN Joint Project ``Ab initio investigation in nuclear physics'',
``Priority Issue on Post-K computer'' (Elucidation of the Fundamental Laws and Evolution of the Universe),
``Program for Promoting Researches on the Supercomputer Fugaku'' (Simulation for basic science: from fundamental laws of particles to creation of nuclei)
and Joint Institute for Computational Fundamental Science (JICFuS).

%%%%%%%%%%%%%%%%%%%%%%%%%%%%%%%%%%%%%%%%%%%%%%%%%%%%%%%%%%%%%%%%%%%%%%%%%%%%%%

%%%%%%%%%%%%%%%%%%%%%%%%%%%%%%%%%%%%%%%%%%%%%%%%%%%%%%%%%%%%%%%%%%%%%%%%%%%%%%

\begin{thebibliography}{99}


\bibitem{Luscher:1990ux} 
  M.~L\"uscher,
  %``Two particle states on a torus and their relation to the scattering matrix,''
  Nucl.\ Phys.\ B {\bf 354}, 531 (1991).
%  doi:10.1016/0550-3213(91)90366-6.

%%%%%%%%%%%%%%%%%%%%%%%%%%%%%%%%%%%%%%%
  
\bibitem{Ishii:2006ec}
N.~Ishii, S.~Aoki and T.~Hatsuda,
%``The Nuclear Force from Lattice QCD,''
Phys. Rev. Lett. \textbf{99}, 022001 (2007)
%doi:10.1103/PhysRevLett.99.022001
[arXiv:nucl-th/0611096 [nucl-th]].

\bibitem{Ishii:2012ssm}
N.~Ishii \textit{et al.} [HAL QCD Coll.],
%``Hadron\textendash{}hadron interactions from imaginary-time Nambu\textendash{}Bethe\textendash{}Salpeter wave function on the lattice,''
Phys. Lett. B \textbf{712}, 437 (2012)
%doi:10.1016/j.physletb.2012.04.076
[arXiv:1203.3642 [hep-lat]].

\bibitem{Aoki:2020bew}
Reviewed in
S.~Aoki and T.~Doi,
%``Lattice QCD and baryon-baryon interactions: HAL QCD method,''
Front. in Phys. \textbf{8}, 307 (2020)
%doi:10.3389/fphy.2020.00307
[arXiv:2003.10730 [hep-lat]].
%and references therein.

%%%%%%%%%%%%%%%%%%%%%%%%%%%%%%%%%%%%%%%

\bibitem{Iritani:2018vfn}
T.~Iritani \textit{et al.} [HAL QCD Coll.],
%``Consistency between L\"uscher\textquoteright{}s finite volume method and HAL QCD method for two-baryon systems in lattice QCD,''
JHEP \textbf{03}, 007 (2019)
%doi:10.1007/JHEP03(2019)007
[arXiv:1812.08539 [hep-lat]].


%%%%%%%%%%%%%%%%%%%%%%%%%%%%%%%%%%%%%%%

\bibitem{Francis:2018qch}
A.~Francis \textit{et al.},
%J.~R.~Green, P.~M.~Junnarkar, C.~Miao, T.~D.~Rae and H.~Wittig,
%``Lattice QCD study of the $H$ dibaryon using hexaquark and two-baryon interpolators,''
Phys. Rev. D \textbf{99}, 074505 (2019)
%doi:10.1103/PhysRevD.99.074505
[arXiv:1805.03966 [hep-lat]].

\bibitem{Horz:2020zvv}
B.~H\"orz \textit{et al.},
%D.~Howarth, E.~Rinaldi, A.~Hanlon, C.~C.~Chang, C.~K\"orber, E.~Berkowitz, J.~Bulava, M.~A.~Clark and W.~T.~Lee, \textit{et al.}
%``Two-nucleon S-wave interactions at the $SU(3)$ flavor-symmetric point with $m_{ud}\simeq m_s^{\rm phys}$: A first lattice QCD calculation with the stochastic Laplacian Heaviside method,''
Phys. Rev. C \textbf{103}, 014003 (2021)
%doi:10.1103/PhysRevC.103.014003
[arXiv:2009.11825 [hep-lat]].

\bibitem{Amarasinghe:2021lqa}
  S.~Amarasinghe \textit{et al.},
  %, R.~Baghdadi, Z.~Davoudi, W.~Detmold, M.~Illa, A.~Parreno, A.~V.~Pochinsky, P.~E.~Shanahan and M.~L.~Wagman,
%``A variational study of two-nucleon systems with lattice QCD,''
[arXiv:2108.10835 [hep-lat]].

%%%%%%%%%%%%%%%%%%%%%%%%%%%%%%%%%%%%%%%

\bibitem{HALQCD:2018gyl}
T.~Iritani \textit{et al.} [HAL QCD Coll.],
%``Systematics of the HAL QCD Potential at Low Energies in Lattice QCD,''
Phys. Rev. D \textbf{99}, 014514 (2019)
%doi:10.1103/PhysRevD.99.014514
[arXiv:1805.02365 [hep-lat]].

%%%%%%%%%%%%%%%%%%%%%%%%%%%%%%%%%%%%%%%

\bibitem{Gongyo:2017fjb}
  S.~Gongyo \textit{et al.} [HAL QCD Coll.],
  %K.~Sasaki, S.~Aoki, T.~Doi, T.~Hatsuda, Y.~Ikeda, T.~Inoue, T.~Iritani, N.~Ishii and T.~Miyamoto, \textit{et al.}
%``Most Strange Dibaryon from Lattice QCD,''
Phys. Rev. Lett. \textbf{120}, 212001 (2018)
%doi:10.1103/PhysRevLett.120.212001
[arXiv:1709.00654 [hep-lat]].

\bibitem{Lyu:2021qsh}
  Y.~Lyu, H.~Tong \textit{et al.},
  %T.~Sugiura, S.~Aoki, T.~Doi, T.~Hatsuda, J.~Meng and T.~Miyamoto,
%``Dibaryon with Highest Charm Number near Unitarity from Lattice QCD,''
Phys. Rev. Lett. \textbf{127}, 072003 (2021)
%doi:10.1103/PhysRevLett.127.072003
[arXiv:2102.00181 [hep-lat]].

%%%%%%%%%%%%%%%%%%%%%%%%%%%%%%%%%%%%%%%

\bibitem{Ishikawa:2015rho}
K.-I.~Ishikawa \textit{et al.} [PACS Coll.],
%``2+1 Flavor QCD Simulation on a $96^4$ Lattice,''
PoS \textbf{LATTICE2015}, 075 (2016)
%doi:10.22323/1.251.0075
[arXiv:1511.09222 [hep-lat]].

\bibitem{Namekawa:2017nfw}
Y.~Namekawa [PACS Coll.],
%``Charm physics by $N_f=2+1$ Iwasaki gauge and the six stout smeared $O(a)$-improved Wilson quark actions on a $96^4$ lattice,''
PoS \textbf{LATTICE2016}, 125 (2017)
doi:10.22323/1.256.0125

%%%%%%%%%%%%%%%%%%%%%%%%%%%%%%%%%%%%%%%

\bibitem{Doi:2012xd}
  T.~Doi and M.~G.~Endres,
  %``Unified contraction algorithm for multi-baryon correlators on the lattice,''
  Comput.\ Phys.\ Commun.\  {\bf 184} (2013) 117
  %doi:10.1016/j.cpc.2012.09.004
  [arXiv:1205.0585 [hep-lat]].


%%%%%%%%%%%%%%%%%%%%%%%%%%%%%%%%%%%%%%%

\bibitem{Amagasa:2015zwb} 
  T.~Amagasa {\it et al.},
  %``Sharing lattice QCD data over a widely distributed file system,''
  J.\ Phys.\ Conf.\ Ser.\  {\bf 664},  042058 (2015).

%%% domain-decomposed solver %%%%
%
%  
%
\bibitem{dd-solver}
%\cite{Boku:2012zi}
%\bibitem{Boku:2012zi}
  T.~Boku {\it et al.},
  %``Multi-block/multi-core SSOR preconditioner for the QCD quark solver for K computer,''
  PoS LATTICE {\bf 2012} (2012) 188
%  doi:10.22323/1.164.0188
  [arXiv:1210.7398 [hep-lat]];
  %%CITATION = doi:10.22323/1.164.0188;%%
%
%
%\bibitem{Terai:2013}
M.~Terai  {\it et al.},
%, K.~I.~Ishikawa, Y.~Sugisaki, K.~Minami, F.~Shoji, Y.~Nakamura, Y.~Kuramashi, M.~Yokokawa,
%"Performance Tuning of a Lattice QCD code on a node of the K computer,"
IPSJ Transactions on Advanced Computing Systems, Vol.6 No.3 43-57 (Sep. 2013);
%(in Japanese);
%
%
%\cite{Nakamura:2011my}
%\bibitem{Nakamura:2011my}
Y.~Nakamura {\it et al.},
%K.-I.~Ishikawa, Y.~Kuramashi, T.~Sakurai and H.~Tadano,
  %``Modified Block BiCGSTAB for Lattice QCD,''
  Comput.\ Phys.\ Commun.\  {\bf 183} (2012) 34
 % doi:10.1016/j.cpc.2011.08.010
  [arXiv:1104.0737 [hep-lat]];
  %%CITATION = doi:10.1016/j.cpc.2011.08.010;%%
%
%  
%\bibitem{Osaki:2010vj}
  Y.~Osaki and K.-I.~Ishikawa,
  %``Domain Decomposition method on GPU cluster,''
  PoS LATTICE {\bf 2010} (2010) 036
%  doi:10.22323/1.105.0036
  [arXiv:1011.3318 [hep-lat]].
  %%CITATION = doi:10.22323/1.105.0036;%%

  
\bibitem{Schrock:2012fj}
  M.~Schr\"ock and H.~Vogt,
  %``Coulomb, Landau and Maximally Abelian Gauge Fixing in Lattice QCD with Multi-GPUs,''
  Comput.\ Phys.\ Commun.\  {\bf 184} (2013) 1907
  [arXiv:1212.5221 [hep-lat]].

\bibitem{bridge}
  \url{https://bridge.kek.jp/Lattice-code/index_e.html}

  
%%%%%%%%%%%%%%%%%%%%%%%%%%%%%%%%%%%%%%%

\end{thebibliography}
\end{document}